\documentclass[11pt,a4paper]{article}
\usepackage{jcappub}
\usepackage[utf8]{inputenc}\usepackage[T1]{fontenc} 
\usepackage{amsmath}
\usepackage{amssymb}
\usepackage{graphicx}
\usepackage{slashed}
\usepackage{cases}
\usepackage[toc,page]{appendix}
\usepackage{empheq}
\usepackage[normalem]{ulem} % to use sout
\usepackage{xcolor} % to use more colors
\usepackage{afterpage}
\usepackage{float}
\usepackage{bbold}
\usepackage{hypbmsec}

\allowdisplaybreaks
\usepackage[export]{adjustbox}

\makeatletter
\newcommand\fake@math{}% just for safety
\def\fake@math#1\){[math]}
\makeatother

\newcommand{\bea}{\begin{eqnarray}}
\newcommand{\eea}{\end{eqnarray}}
\begin{document}

\title{Collider Tests of Flavored Resonant Leptogenesis in the $U(1)_X$ Model}
\author[]{Garv Chauhan}
\affiliation[]{Center for Neutrino Physics, Department of Physics, Virginia Tech, Blacksburg, VA 24061, USA}
\emailAdd{gchauhan@vt.edu}
 % \emailAdd{adas@particle.sci.hokudai.ac.jp}
 % \emailAdd{bdev@wustl.edu}
%%%%%%%%%%%%%%%%%%%%%%%%%%%%%%%%%%%%%%%%%
% For RevTex
% \begin{abstract}
% \end{abstract}

% For JHEP
\abstract{We study the generation of baryon asymmetry through the flavored resonant leptogenesis in the $U(1)_X$ extension of the Standard Model. Being a generalization of the $U(1)_{B\text{-}L}$, $U(1)_X$ is an ultraviolet-complete model of the right-handed neutrinos (RHNs), whose CP violating out-of-equilibrium decays lead to the generation of baryon asymmetry via leptogenesis. We can also explain the neutrino masses via the seesaw mechanism in this model. We consider three different cases for different $U(1)_X$ charges of the scalar particle responsible for $U(1)_X$ breaking at TeV-scale. These include the popular $U(1)_{B\text{-}L}$ and $U(1)_{R}$ models, as well as a $U(1)_C$ model which maximizes the collider signal. We numerically solve the flavored Boltzmann transport equations to calculate the total baryon asymmetry. We show that all three cases considered here can naturally explain the observed baryon asymmetry of the Universe in a large portion of the available parameter space, while satisfying the neutrino oscillation data. We find that the $U(1)_C$ case offers successful leptogenesis in a larger portion of the parameter space as compared to $U(1)_{B\text{-}L}$ and $U(1)_{R}$. We also perform a comparative study between the flavored and unflavored leptogenesis parameter space. Finally, we also study the collider prospects for all these scenarios using the lepton number violating signal of $pp\to \ell^\pm \ell^\pm+$jets mediated by the $Z'$ boson associated with $U(1)_X$. We find that HL-LHC may be able to probe a small portion of the relevant parameter space having successful leptogenesis, if neutrinos have normal mass ordering, while a $\sqrt s=100$ TeV future collider can access a much larger region of the parameter space, thereby offering an opportunity to test resonant leptogenesis in the $U(1)_X$ model.}
%%%%%%%%%%%%%%%%%%%%%%%%%%%%%%%%%%%%%%%%%

% \keywords{Discrete Symmetries, Beyond Standard Model, Neutrino Physics}

\maketitle

%%%%%%%%%%%%%%%%%%%%%
\section{Introduction}
%%%%%%%%%%%%%%%%%%%%%%%%%
Although the Standard Model (SM) of particle physics has been highly successful in describing the microscopic physics to the smallest length scales probed so far, there is strong empirical evidence for physics beyond the SM (BSM). In particular, the observation of neutrino oscillations~\cite{ParticleDataGroup:2022pth} necessarily implies that at least two of the three neutrino mass eigenvalues must be nonzero, which immediately demands some BSM explanation, as neutrinos are exactly massless in the SM. Perhaps the simplest way to generate neutrino mass is by the so-called seesaw mechanism~\cite{Minkowski:1977sc,Yanagida:1979as,Gell-Mann:1979vob,Mohapatra:1979ia,Glashow:1979nm} where one adds SM-singlet Majorana fermions, also known as right-handed neutrinos (RHNs), which give rise to a small Majorana mass for the $SU(2)_L$-doublet neutrinos after electroweak symmetry breaking. 

It is interesting that the same RHNs responsible for neutrino mass could also explain another important evidence for BSM physics, namely, the observed matter-antimatter asymmetry of the Universe~\cite{Planck:2018vyg}, via the mechanism of leptogenesis~\cite{Fukugita:1986hr}. The basic idea is that the out-of-equilibrium decays of the RHNs to the SM lepton and Higgs doublets can produce a nonzero lepton asymmetry in the early Universe, which is reprocessed into a baryon asymmetry by the $(B+L)$-violating electroweak sphaleron transitions~\cite{Kuzmin:1985mm}. However, the vanilla leptogenesis scenario imposes a lower bound on the mass of the RHNs, $M_N\gtrsim 10^9$ GeV~\cite{Davidson:2002qv, Buchmuller:2002rq},\footnote{Including flavor effects can in principle lower this value to $\sim 10^6$ GeV~\cite{Moffat:2018wke}.} thus precluding the possibility of testing it in laboratory experiments. 

A low-energy alternative is the resonant leptogenesis mechanism~\cite{Pilaftsis:2003gt}, which relies on the resonant enhancement of the CP asymmetry from RHN decays via self-energy contributions when the masses of two RHNs are quasi-degenerate~\cite{Pilaftsis:1997jf}. This can potentially bring the scale of $M_N$ down to the electroweak scale\footnote{RHN scale can even be lighter below GeV-scale in the parametric regime of ARS leptogenesis~\cite{Drewes:2017zyw}.}~\cite{Pilaftsis:2005rv, Deppisch:2010fr, BhupalDev:2014pfm, Dev:2017wwc}, thus offering the hope to test this mechanism at the Large Hadron Collider (LHC) and future colliders. 

In this work, we study the resonant leptogenesis mechanism in an ultraviolet-complete model of the RHNs in terms of the $U(1)_X$ extension of the SM~\cite{Appelquist:2002mw,Iso:2010mv,Coriano:2014mpa, Oda:2015gna,Das:2016zue}. The $U(1)_X$ symmetry can be identified as the linear combination of the $U(1)_Y$ in SM and the $U(1)_{B-L}$ gauge group, and hence, can be regarded as a generalization of the $U(1)_{B-L}$ extension of the SM~\cite{Davidson:1978pm, Marshak:1979fm, Buchmuller:1991ce}, where the RHN fields are an essential ingredient required for anomaly cancellation. The presence of the extra neutral gauge boson $Z'$ associated with the $U(1)_X$ breaking affects the lepton asymmetry calculation in a nontrivial way~\cite{Blanchet:2009bu}. We take this into account and also include the flavor effects from both RHNs and charged leptons, which are known to be important in resonant leptogenesis~\cite{BhupalDev:2014pfm, Dev:2017trv}. A crucial input for leptogenesis is the complex Dirac Yukawa coupling matrix, which we parametrize using the Casas-Ibarra parametrization~\cite{Casas:2001sr} to satisfy the neutrino oscillation data, while also highlighting the role of the Dirac CP phase in the generation of the lepton asymmetry. We then perform a numerical scan over the masses of the heavy gauge boson and RHNs to carve out the parameter space consistent with successful leptogenesis for three different benchmark charges of the $U(1)_X$ scalars. 

We also calculate the collider prospects of the allowed parameter space for leptogenesis using the lepton number violating (LNV) signal $pp\to Z'\to NN\to \ell^\pm \ell^\pm+{\rm jets}$~\cite{Buchmuller:1991ce, Basso:2008iv, FileviezPerez:2009hdc, Kang:2015uoc, Cox:2017eme, Chauhan:2021xus}. We find that the high-luminosity LHC (HL-LHC) will be sensitive to a small portion of  the allowed parameter space, especially for the normal hierarchy of neutrino masses, while a future 100 TeV collider can probe a much wider range of parameter space, thus making resonant leptogenesis in the $U(1)_X$ model truly testable at the Energy Frontier.   

% The rest of the paper is organized as follows: In Section~\ref{model} we briefly describe the $U(1)_X$ model. In Section~\ref{lepto}, we analyze the resonant leptogenesis scenario. In Section~\ref{osc} we apply neutrino oscillation data to investigate the role of Dirac CP phase to generate the right amount of baryon asymmetry of the Universe. We conclude the article in Sec.~\ref{conc}.
%%%%%%%%%%%%%%%%%%%%%%%%%%%%%%%%%%%%%%%%%%%%%%%%%%%
\section{General $U(1)_X$ Scenario}
\label{sec:model}
%%%%%%%%%%%%%%%%%%%%%%%%%%%%%%%%%%%%%%%%%%%%%%%%%%
The general $U(1)_X$ extension of the SM is based on the $SU(2)_C \times SU(2)_L \times U(1)_Y \times U(1)_X$ gauge group. The particle content involves adding three generations of the SM-singlet RHNs and a SM-singlet scalar $\Phi$, all charged under $U(1)_X$. The RHNs in addition to contributing to the neutrino mass generation via seesaw mechanism, also play a crucial role in canceling the gauge and mixed gauge-gravity anomalies~\cite{Das:2016zue}. The particle content of the model along with the $U(1)_X$ charges is given in Table~\ref{tab1}. Note that the fermion charges are generation-independent. The scalar charges $x_H$, $x_\Phi$ are real parameters. For $x_H=0$ and $x_\Phi=1$, we recover the $U(1)_{B-L}$ model~\cite{Davidson:1978pm, Marshak:1979fm}. Without loss of generality, we fix $x_\Phi=1$ in this paper. As a result, the $x_H$ charge simply acts as an angle between the $U(1)_Y$ and $U(1)_{B-L}$ directions. $x_H=-2$ corresponds to the $U(1)_R$ model~\cite{Dutta:2019fxn}, whereas $x_H=-1.2$ corresponds to the $U(1)_C$ model with maximum enhancement in the $Z'\to NN$ branching ratio~\cite{Das:2017flq}. We will consider these three benchmark values of $x_H$ in the following numerical analysis. The $U(1)_X$ gauge coupling $g_X$ is another free parameter in this model and we will also fix some benchmark values for it for a given $Z'$ mass to be consistent with the current LHC constraints~\cite{Das:2021esm}. 

The fermion mass terms and flavor mixing are introduced by the Yukawa interaction terms written as 
\begin{equation}
{\cal L}_{\rm Yukawa} = 
				 - Y_e^{\alpha \beta} \overline{\ell_L^\alpha} \tilde{H} e_R^\beta
				- Y_\nu^{\alpha \beta} \overline{\ell_L^\alpha} H N_R^\beta- Y_N^\alpha \Phi \overline{(N_R^\alpha)^c} N_R^\alpha - Y_u^{\alpha \beta} \overline{q_L^\alpha} H u_R^\beta
                 - Y_d^{\alpha \beta} \overline{q_L^\alpha} \tilde{H} d_R^\beta
    +  {\rm H.c.}~,
\label{LYk}   
\end{equation}
where $\tilde{H} = i  \sigma^2 H^*$ and $\alpha, \beta=1,2,3$ are the generation indices. Note here $H$ is the SM Higgs doublet. %%%%%%%%%%%%%%%%%%%%%%%%%%%%%%%%%%%%%%%%%%%%%
\begin{table}[t!]
\begin{center}
\begin{tabular}{|c|ccc||c|}
\hline
            & $SU(3)_c$ & $SU(2)_L$ & $U(1)_Y$ & $U(1)_X$ \\[2pt]
\hline
$q_L^\alpha$    & {\bf 3}   & {\bf 2}& $\frac{1}{6}$ & 		 $q_q=\frac{1}{6}x_H + \frac{1}{3}x_\Phi$   \\[2pt] 
$u_R^\alpha$    & {\bf 3} & {\bf 1}& $\frac{2}{3}$ &  	  $q_u=\frac{2}{3}x_H + \frac{1}{3}x_\Phi$ \\[2pt] 
$d_R^\alpha$    & {\bf 3} & {\bf 1}& $-\frac{1}{3}$ & 	 $q_d=-\frac{1}{3}x_H + \frac{1}{3}x_\Phi$  \\[2pt] 
\hline
$\ell_L^\alpha$    & {\bf 1} & {\bf 2}& $-\frac{1}{2}$ & 	 $q_\ell=- \frac{1}{2}x_H - x_\Phi$  \\[2pt] 
$e_R^\alpha$   & {\bf 1} & {\bf 1}& $-1$   &		$q_e=- x_H - x_\Phi$  \\[2pt] 
\hline
$N_\alpha$   & {\bf 1} & {\bf 1}& $0$   &	 $q_N=- x_\Phi$ \\[2pt] 
\hline
$H$         & {\bf 1} & {\bf 2}& $-\frac{1}{2}$  &  	 $\frac{x_H}{2}$  \\ 
$\Phi$      & {\bf 1} & {\bf 1}& $0$  & 	 $2 x_\Phi$   \\
\hline
\end{tabular}
\end{center}
\caption{
The particle content of  the general $U(1)_X$ model. Here $\alpha=1, 2, 3$ represents the family index. }
\label{tab1}
\end{table}
%%%%%%%%%%%%%%%%%%%%%%%%%%%%%%%%%%%%%%%%%%%%%%%%

The renormalizable scalar potential in this model is given by
\begin{align}
  V \ = \ m_H^2(H^\dag H) + \lambda_H^{} (H^\dag H)^2 + m_\Phi^2 (\Phi^\dag \Phi) + \lambda_\Phi^{} (\Phi^\dag \Phi)^2 + \lambda_{\rm mix} (H^\dag H)(\Phi^\dag \Phi)~.
\end{align}
In the limit of small $\lambda_{\rm mix}$, the scalar fields $H$ and $\Phi$ can be analyzed separately~\cite{Oda:2015gna,Das:2016zue}. The $U(1)_X$ gauge symmetry and electroweak symmetry are respectively broken by the vacuum expectation values (VEVs) of $\Phi$ and $H$, given by 
\begin{equation}
  \langle H\rangle  \ = \ \frac{1}{\sqrt{2}}\begin{pmatrix} v+h\\0 
  \end{pmatrix}, \quad {\rm and}\quad 
  \langle\Phi\rangle \ =\  \frac{v_\Phi^{}+\phi}{\sqrt{2}},
\end{equation}
where $v=246$ GeV is the electroweak scale and $v_\Phi$ is a free parameter, which can be traded for the $Z'$ mass. In other words, after the $U(1)_X$ symmetry is broken and assuming $v_\Phi \gg v$, the $Z^\prime$ mass can be written as 
\begin{equation}
 m_{Z^\prime} \simeq  2 g_X  v_\Phi x_\Phi \, .
\end{equation}
It can be clearly seen after the breaking of $U(1)_X$ and $\Phi$ VEV is developed, the third term in the Eq.~\eqref{LYk}
leads to the generation of Majorana mass of the RHNs. The second term after the $H$ develops VEV generates the Dirac mass term from the Yukawa coupling. The combination of these Dirac and Majorana mass terms leads to Type-I seesaw formula that can explain the neutrino masses and mixing. The  Dirac and Majorana mass terms arising from Eq.~\eqref{LYk} can be written as
\begin{equation}
    m_{D}^{\alpha \beta} \  =  \ \frac{Y_{\nu}^{\alpha \beta}}{\sqrt{2}} v, \, \, \, \, \,
    m_{N^\alpha}^{} \ = \ \frac{Y^\alpha_{N}}{\sqrt{2}} v_\Phi^{}.
\label{mDI}
\end{equation}
respectively. 
%%%%%%%%%%%%%%%%%%%%%%%%%%%%%%%%%%%%%%%%%%%%%%%%%%%%%%%%
\subsection{Fermions and $Z^\prime$ interactions}
%%%%%%%%%%%%%%%%%%%%%%%%%%%%%%%%%%%%%%%%%%%%%%%%%%%%%%%

The new gauge boson $Z'$ interacts with the SM fermions via $U(1)_X$ gauge coupling. The chiral interaction terms in the  Lagrangian for fermions interacting with $Z'$ is given by
\begin{eqnarray}
\mathcal{L}_{\rm{int}} = -g_X (\overline{f}\gamma_\mu q_{f_{L}^{}}^{} P_L^{} f+ \overline{f}\gamma_\mu q_{f_{R}^{}}^{}  P_R^{} f) Z_\mu^\prime~,
\label{Lag1}
\end{eqnarray}
where $q_{f_{L(R)}}$ is the corresponding $U(1)_X$ charge of the left (right) handed fermions [cf.~Table~\ref{tab1}] and $P_{L(R)}^{}= (1 \pm \gamma_5)/2$  are the usual projection operators. Using this, we can calculate the partial decay widths of $Z^\prime$ into charged fermions as follows 
\begin{align}
\label{eq:width-ll}
    \Gamma(Z' \to \bar{f} f)
    &= N_C^{} \frac{M_{Z'}^{} g_{X}^2}{24 \pi} \sqrt{1 - 4\frac{m_f^2}{M_{Z'}^2}}\left[ \left( q_{f_L^{}}^2 + q_{f_R^{}}^2 \right) \left( 1 - \frac{m_f^2}{M_{Z'}^2} \right) + 6 q_{f_L^{}}^{} q_{f_R^{}}^{} \frac{m_f^2}{M_{Z'}^2} \right]~,
\end{align}    
% where $m_f$ is the SM fermion mass and $N_C^{}=1~(3)$ for the SM leptons (quarks). The partial decay width of $Z^\prime$ into a pair of light neutrinos for three generations neglecting neutrino masses is given by
% \begin{align}   
% \label{eq:width-nunu}
%     \Gamma(Z' \to \nu \nu)
%     = 3 \frac{M_{Z'}^{} g_{X}^2}{24 \pi} q_{f_L^{}}^2~,
% \end{align} 
% where $q_{f_L^{}}^{}$ is the $U(1)_X$ charge of the SM lepton doublet, 
For our collider study, the relevant decay mode is $Z^\prime\to NN$ whose partical decay width is given by
\begin{align}
\label{eq:width-NN}
    \Gamma(Z' \to N_\alpha N_\alpha)
    = \frac{M_{Z'}^{} g_{X}^2}{24 \pi} q_{N^{}}^2 \left( 1 - \frac{M_N^2}{M_{Z'}^2} \right)^{\frac{3}{2}}~,
\end{align}
where $q_N$ is the $U(1)_X$ charge of RHNs and $M_N$ is the RHN mass. 
%%%%%%%%%%%%%%%%
\subsection{RHN interactions}
\label{HNL}
%%%%%%%%%%%%%%%%%%%%%%%%%%%%
As discussed earlier after breaking of $U(1)_X$ and EW symmetry i.e. after $\Phi$ and $H$ develop VEVs, it leads to the generation of RHN Majorana mass $M_N$ and Dirac mass term respectively. This leads to the generation of neutrino masses in this model. The full neutrino mass matrix takes the standard seesaw form as  
%%%%%%%%%%%%%%%
\begin{eqnarray}
M_{\nu} \ = \ \begin{pmatrix}
0&&M_{D}\\
M_{D}^{T}&&M_{N}
\end{pmatrix},  
\label{typeInu}
\end{eqnarray}
where we can consider $M_N$ as a diagonal matrix without the loss of generality.  Diagonalizing this mass matrix we obtain the light neutrino mass eigenvalues as
\bea
m_\nu \simeq -M_D M_N^{-1} M_D^T, 
\eea
in the seesaw limit $m_D\ll M_N$.  

The light neutrino flavor eigenstate $(\nu_\alpha)$ can be approximately decomposed in terms of light $(\nu_i)$ and heavy $(N_i)$ mass eigenstates
\bea 
\nu_\alpha \simeq  U_{\alpha i}\nu_i  + V_{\alpha i} N_i,  
\eea 
where $\alpha$ and $i$ are the generation indices. Note here $U_{\alpha i}$ are the elements of the $3 \times 3$ light neutrino mixing matrix which can be expressed as $(1-\frac{\epsilon}{2})U_{\rm PMNS}$ with $\epsilon= V^\ast V^T$ known as the non-unitarity parameter, with 
\bea
V_{\alpha i} \simeq M_D M_N^{-1}
\label{eq:mixing}
\eea
parametrizing the mixing between light and RHNs. Therefore, the SM gauge singlet RHNs interact with the SM gauge bosons through this light-heavy mixing. The light neutrino mass matrix can be diagonalized by a $3\times3$ matrix here denoted as $U_{\rm PMNS}$
\bea
U_{\rm PMNS}^T~m_\nu~U_{\rm PMNS} = {\rm diag}(m_1, m_2, m_3).
\eea 
As expected, a non-zero $\epsilon$ can lead to the mixing matrix $U$ becoming non-unitary. But given the stringent constraints on $\epsilon$~\cite{Blennow:2023mqx}, we can take $U\simeq U_{\rm PMNS}$ without affecting our leptogenesis results. 

Due to the light-RHN mixing, the charged-current interactions can be expressed in terms of neutrino mass eigenstates as
\bea 
{\mathcal{L}_{\rm CC} \supset 
 -\frac{g}{\sqrt{2}} W_{\mu}
  \bar{e}_\alpha \gamma^{\mu} P_L   V_{\alpha i} N_i  + \rm{h.c}.} 
\label{CC}
\eea
Similarly, the neutral-current interactions in terms of the mass eigenstates can be written as
\bea 
{\mathcal{L}_{\rm NC} \supset 
 -\frac{g}{2 \cos \theta_w}  Z_{\mu} 
\left[ \left( 
  \overline{\nu}_m \gamma^{\mu} P_L (U^{\dagger}V)_{mi}  N_i
  + \rm{h.c.} \right) +
  \overline{N}_m \gamma^{\mu} P_L  (V^{\dagger} V)_{mi} N_i
\right] , }
\label{NC}
\eea
where  $\theta_w$ is the weak mixing angle. Due to these interactions, the RHNs can decay into $\ell W$, $\nu Z$ and $\nu h$ final states. For RHNs heavier than $W$, $Z$ and $h$, the decays are characterized as being 2 body on-shell decays, where SM bosons will decay further into lighter SM particles. The partial decay widths for these three processes are 
\bea
\Gamma(N_i \rightarrow \ell_{\alpha} W)
 &=& \frac{|V_{\alpha i}|^{2}}{16 \pi} 
\frac{ (M_{N_i}^2 - M_W^2)^2 (M_{N_i}^2+2 M_W^2)}{M_{N_i}^3 v_h^2} ,
\nonumber \\
\Gamma(N_i \rightarrow \nu_\alpha Z)
&=& \frac{|V_{\alpha i}|^{2}}{32 \pi} 
\frac{ (M_{N_i}^2 - M_Z^2)^2 (M_{N_i}^2+2 M_Z^2)}{M_{N_i}^3 v_h^2} ,
\nonumber \\
\Gamma(N_i \rightarrow \nu_\alpha h)
 &=& \frac{|V_{\alpha i}|^{2}}{32 \pi}\frac{(M_{N_i}^2-M_h^2)^2}{M_{N_i} v_h^2}.
\label{eq:dwofshell}
\eea 
We will use the decay branching ratios of the RHNs into $\ell_\alpha W$ in our collider analysis. 
%%%%%%%%%%%%%%%%%%%%%%%%%%%%%%%%%%%%%%%%%%%%%%%%%%%%%%%
\section{Resonant leptogenesis in general $U(1)_X$ scenario}
\label{lepto}

The CP violating decays of the right-handed neutrinos are responsible for the generation of lepton asymmetry in the resonant leptogenesis mechanism. The amount of flavored asymmetry generated is proportional to the CP asymmetry ($\epsilon_{CP}$) in these decays. 
\begin{align}
    \varepsilon_{i\alpha} \, = \, \frac{\Gamma(N_i\to L_\alpha H)-\Gamma(N_i\to \bar{L}_\alpha H^c)}{\Gamma(N_i\to L_\alpha H)+\Gamma(N_i\to \bar{L}_\alpha H^c)} \, ,
    \label{eq:eps1}
\end{align}
The $(B+L)$-violating electroweak sphaleron processes convert this lepton asymmetry at $T_c$ to baryon asymmetry, where below the critical temperature $T_c\sim 149$ GeV these sphaleron processes freeze-out. 

The dominant contribution to the lepton asymmetry arises from the interference between the tree and self-energy diagrams in the $N_i$ decay. This contribution can be enhanced if the intermediate state $N_j(j\neq i)$ is quasi-degenerate with $N_i$. This resonantly enhanced $\varepsilon_{CP}$ in terms of RHN masses and neutrino Dirac Yukawa matrix $Y_D$ can be expressed as :
\begin{align}
    \varepsilon_{i\alpha}  \simeq  \frac{1}{8\pi \left(Y_D^\dag Y_D\right)_{ii}} \sum_{j\neq i} {\rm Im}\left[\left(Y_D^\star\right)_{\alpha i} \left(Y_D\right)_{\alpha j}\right] \, {\rm Re}\left[\left(Y_D^\dag Y_D\right)_{ij}\right]\frac{M_i M_j (M_i^2-M_j^2)}{(M_i^2-M_j^2)^2+A_{ij}^2}  \, , 
    \label{eq:eps2}
\end{align}
where $A_{ij}$ denotes the regulator controlling the behaviour of decay asymmetry in the degenerate limit $\Delta M_{ij}\rightarrow 0$. There are two different contributions to total CP asymmetry from RHN mixing as well as from oscillations, essentially with a similar form as above but with different regulators. 
\begin{align}
    A_{ij}^{\rm mix} \, &= \, M_i \Gamma_j \, , \quad \nonumber \\ \quad A_{ij}^{\rm osc} \, &= \, (M_1\Gamma_1+M_2\Gamma_2)\left[\frac{{\rm det}\left({\rm Re}\left(Y_D^\dag Y_D\right)\right)}{\left(Y_D^\dag Y_D\right)_{ii}\left(Y_D^\dag Y_D\right)_{jj}} \right]^{1/2} \, .
    \label{eq:regulator}
\end{align}
Therefore, the total CP asymmetry is given by $\varepsilon_{\rm tot}=\varepsilon_{\rm mix}+ \varepsilon_{\rm osc}$~\cite{BhupalDev:2014pfm, BhupalDev:2014oar, Klaric:2020phc}.

After semi-analytically solving the flavored Boltzmann transport equations, the total lepton asymmetry produced after the sphaleron freezeout can be written in the following form
\begin{align}
    \eta_{L_\alpha} \ \simeq \ \frac{3}{2z_cK_\alpha^{\rm eff}}\sum_i \varepsilon_{i\alpha}d_i \, ,
    \label{eq:etaL}
\end{align}
where $z_c=M_N/T_c$, $K_\alpha^{\rm eff}$ are the effective washout factors in presence of $Y_D$ and any additional interactions (including the effect of the real intermediate state subtracted collision terms), and $d_i$ are the corresponding dilution factors given in terms of ratios of thermally-averaged rates for decays and scatterings involving $N_i$ (see Ref.~\cite{BhupalDev:2014pfm} for details).

The total baryon asymmetry $\eta_B$ generated from $\eta_{L}$ after taking  sphaleron efficiency and entropy dilution into account is given by 
\begin{equation}
    \eta_{B} = -\frac{28}{79} \frac{1}{27.3} \sum_\alpha \eta_{L_\alpha}
\end{equation}
This is to be compared against the observed baryon asymmetry of the Universe $\eta_{\text{BAU}} = (6.12 \pm 0.08) \times 10^{-10}$~\cite{Planck:2018vyg}.

Since enough $\eta_B$ comparable to the $\eta_{\text{BAU}}$ can be produced in multiple corners of the parameter space of the $U(1)_X$ model, we specifically focus on maximizing the $\eta_B$ produced for a given RHN mass scale $M_N$ and $U(1)_X$ gauge boson mass $M_Z'$. Given the time-complexity of the flavored Boltzmann equations, in general maximizing $\eta_B$ is a formidable task. Therefore, we choose to maximize CP asymmetry parameter $\varepsilon_{tot}$ as a function of $Y_D$ and $\Delta M_N$. We can conveniently parameterize $Y_D$ using the Casas-Ibarra form 
\begin{equation}
    Y_D \, = \, \frac{\sqrt{2}\,i}{v}U_{\text{PMNS}}\sqrt{{D}_\nu}O\sqrt{{D}_N} \, 
\end{equation}
where $D_N = \text{diag}(M_1,M_2)$, $D_\nu = \text{diag}(m_1,m_2,m_3)$\footnote{To reduce the number of free parameters in this study, we choose the lightest neutrino to be massless.} and $O$ is an arbitrary $2\times 3$ orthogonal matrix. When the total CP asymmetry contribution is dominated by the mixing case, the $\Delta M_N$ that maximizes $\varepsilon$ is given by $ \Delta M_N \sim 0.5 \Gamma_N $, where $\Gamma_N$ is the average decay width of $N_i$-pair. However for low-scale leptogenesis both contributions are of equal importance, which leads to a modified relation for the \textit{optimum} mass-splitting~\cite{Chauhan:2021xus}
\begin{equation}
   {\Delta\, M_{N,\,{\rm opt}}} \, \sim 1.23\, {\Gamma_N} \, ,
   \label{eq:kappamax}
\end{equation}
The above factor of 1.23 is obtained numerically by maximizing $\varepsilon_{\rm tot}$ including the contributions from both regulators $A_{ij}^{\rm mix}$ and $A_{ij}^{\rm osc}$. Since $\Gamma_N$ scales as $M_N^2$ (also including the $M_N$ dependence of $Y_D$), $\Delta\, M_{N,\,{\rm opt}}$ increases quadratically with increasing $M_N$. 

%%%%%%%%%%%%%%%%%%%%%%%%%%%%%%%%%%%%%%%%%%%%%%%%%%%
% \begin{figure}
%    \centering
%     \includegraphics[width=0.49\textwidth]{Lep_NH_gBL1_LHC_100TeV.pdf}
%     \includegraphics[width=0.49\textwidth]{Lep_NH_gR1_LHC_100TeV.pdf}
%     \caption{Caption}
%     \label{fig:my_label-1}
% \end{figure}
% \begin{figure}
%    \centering
%     \includegraphics[width=0.49\textwidth]{Lep_IH_gBLmixed_LHC_100TeV.pdf}
%     \includegraphics[width=0.49\textwidth]{Lep_IH_gRmixed_LHC_100TeV.pdf}
%     \caption{Caption}
%     \label{fig:my_label-2}
% \end{figure}
% \begin{figure}
%    \centering
%     \includegraphics[width=0.65\textwidth]{Lep_NH_gBL1_mixedDeltaMN_LHC_100TeV.pdf}
%     \caption{Caption}
%     \label{fig:my_label-2}
% \end{figure}

\begin{figure}
   \centering
    \includegraphics[width=0.49\textwidth]{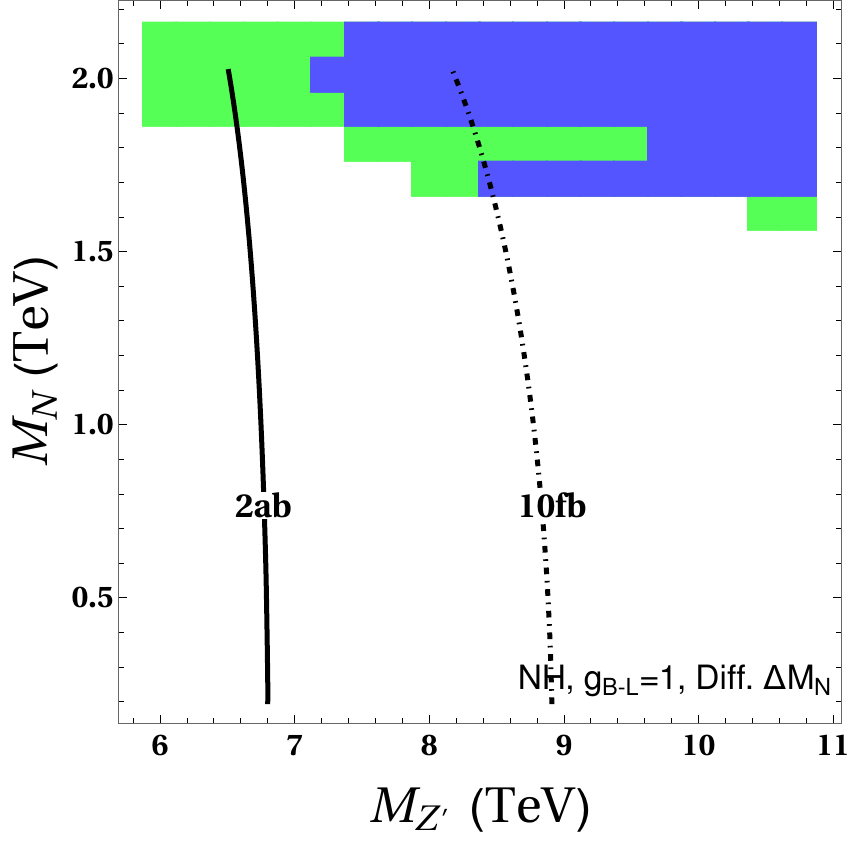}
    \includegraphics[width=0.49\textwidth]{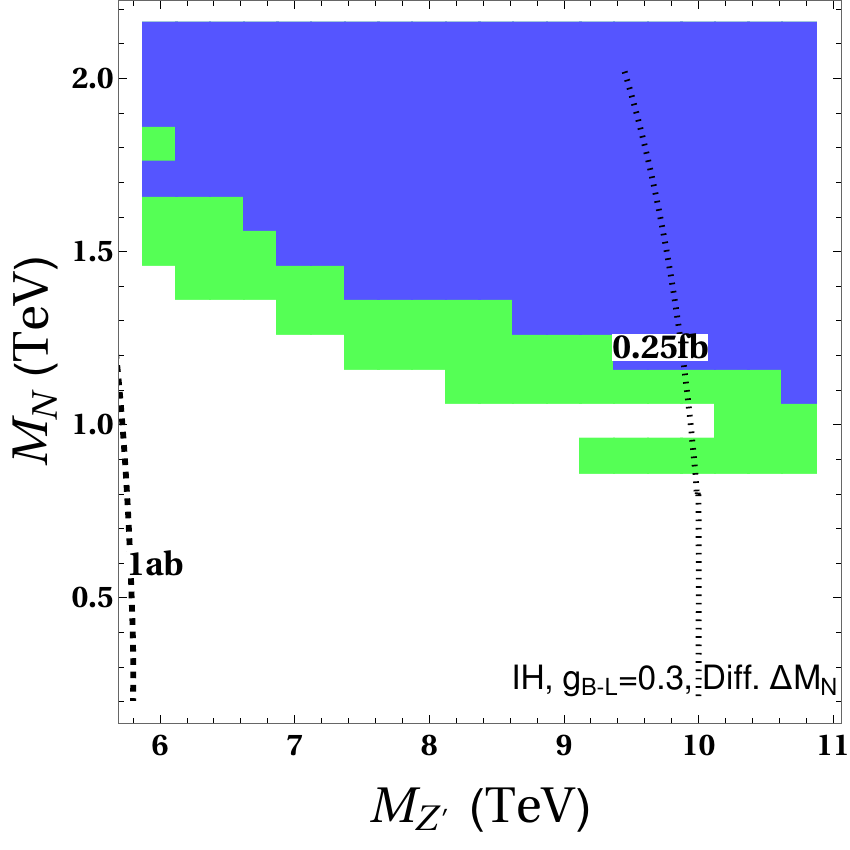}
    \caption{Prediction of the baryon asymmetry for $\eta_B \geq \eta_B^{\rm obs}$ in the $(M_{Z'},M_N)$ plane for a fixed $g_{B-L}=1\,(0.3)$ for NH (IH) in the $U(1)_{B-L}$ case. The green region indicates optimum RHN mass splitting ($\Delta M_{N,\text{opt}}$, see Eq.~\eqref{eq:kappamax}, and blue region indicates $\Delta M_N = 10\,\Delta M_{N,\text{opt}}$. The contours show $\sigma_{\mathrm{prod}}$ (in ab/fb) at the $\sqrt s=14$ TeV LHC (solid/dashed) and at $\sqrt s=100$ TeV future collider (dot-dashed, dotted). Note : Blue region lies completely inside the green region. }
    \label{fig:1BfMN}
\end{figure}

\begin{figure}
   \centering
    \includegraphics[width=0.49\textwidth]{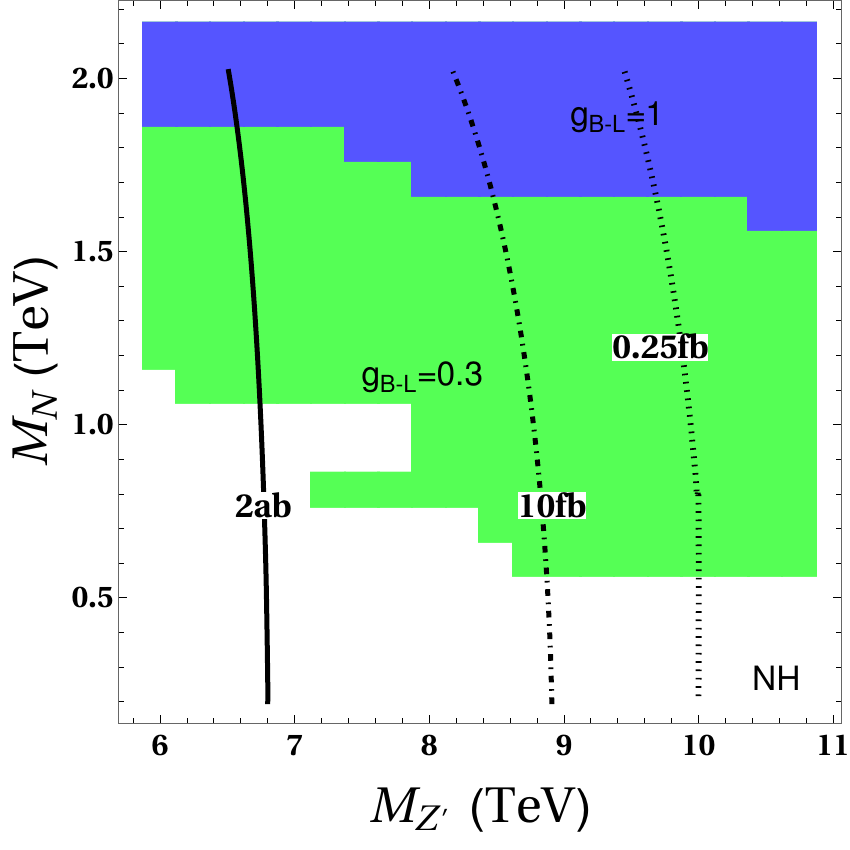}
    \includegraphics[width=0.49\textwidth]{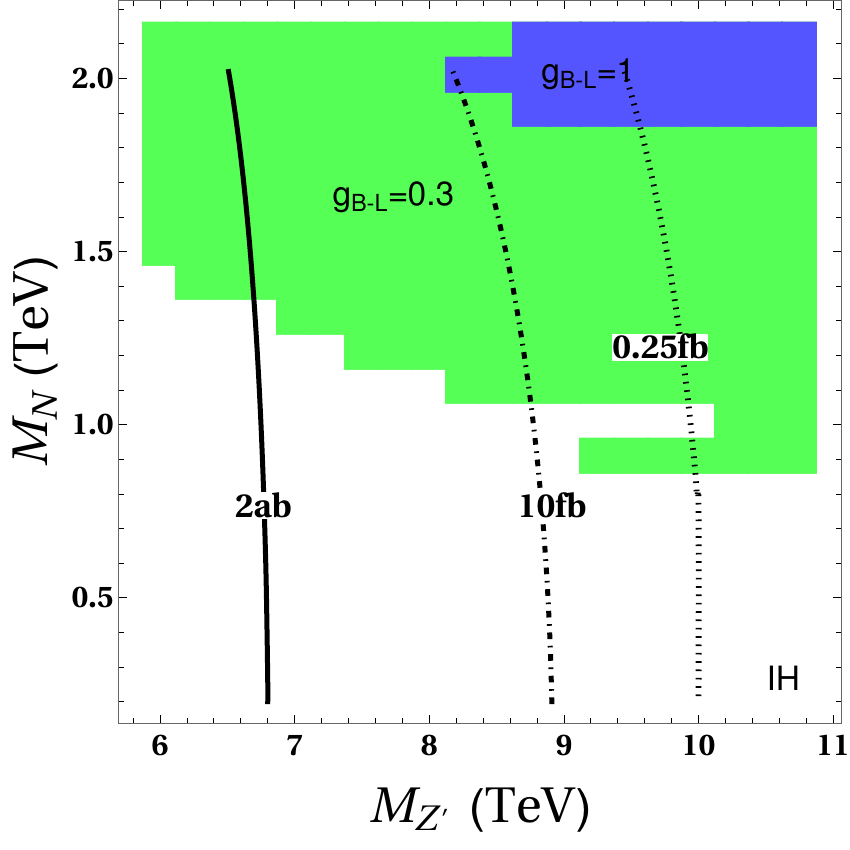}
    \caption{Prediction of the baryon asymmetry for $\eta_B \geq \eta_B^{\rm obs}$ shown in green (blue) region in the $(M_{Z'},M_N)$ plane for $g_{B-L}=0.3\,(1)$ for both NH and IH in $U(1)_{B-L}$ case. The contours show $\sigma_{\mathrm{prod}}$ (in ab/fb) at the $\sqrt s=14$ TeV LHC ($g_{B-L}=1$, solid) and at $\sqrt s=100$ TeV future collider ($g_{B-L}=0.3$, dashed) and ($g_{B-L}=1$, dot-dashed). Note : Blue region lies completely inside the green region.}
    \label{fig:1Bf}
\end{figure}

\begin{figure}
   \centering
   \includegraphics[width=0.49\textwidth]{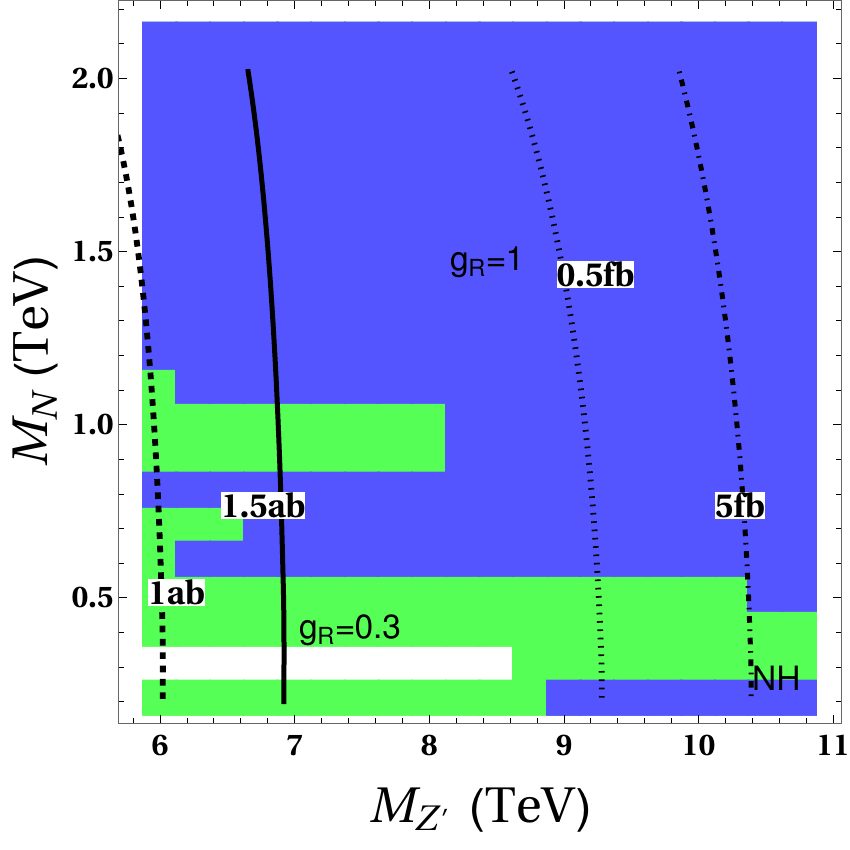}
    \includegraphics[width=0.49\textwidth]{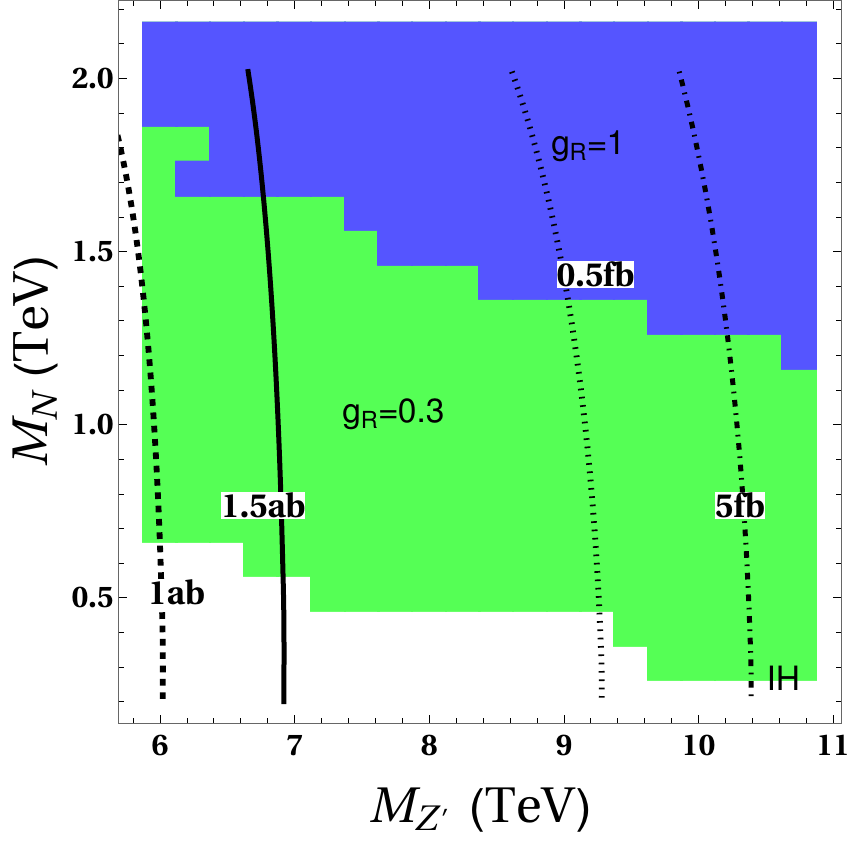}
    \caption{Prediction of the baryon asymmetry for $\eta_B \geq \eta_B^{\rm obs}$ shown in green (blue) region in the $(M_{Z'},M_N)$ plane for $g_{R}=0.3\,(1)$ for both NH and IH in $U(1)_{R}$ case. The contours show $\sigma_{\mathrm{prod}}$ (in ab/fb) at the $\sqrt s=14$ TeV LHC ($g_{R}=1$, solid), ($g_{R}=0.3$, dotted) and at $\sqrt s=100$ TeV future collider ($g_{R}=0.3$, dashed) and ($g_{R}=1$, dot-dashed). Note : Blue region lies completely inside the green region.}
    \label{fig:1Rf}
\end{figure}

\begin{figure}
   \centering
    \includegraphics[width=0.49\textwidth]{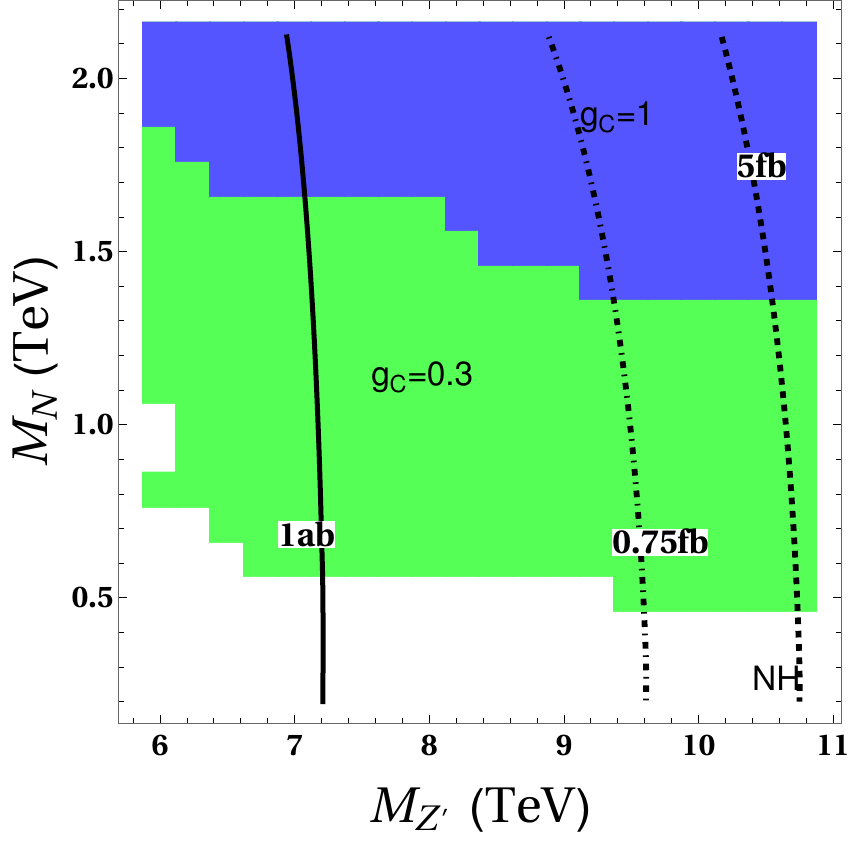}
    \includegraphics[width=0.49\textwidth]{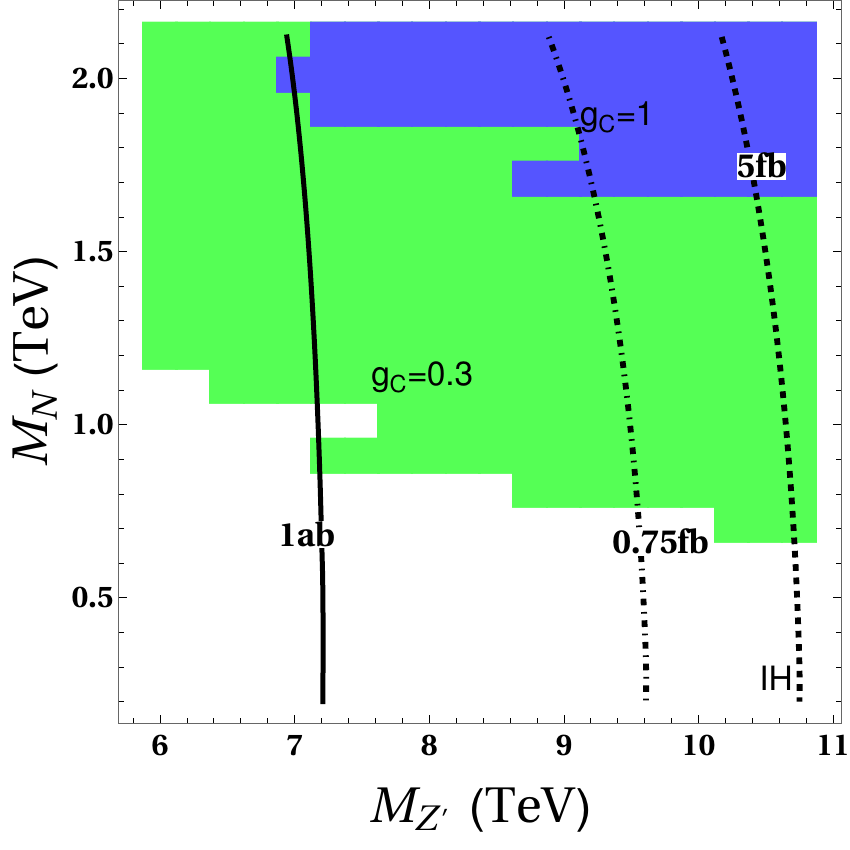}
    \caption{Prediction of the baryon asymmetry for $\eta_B \geq \eta_B^{\rm obs}$ shown in green (blue) region in the $(M_{Z'},M_N)$ plane for $g_{C}=0.3\,(1)$ for both NH and IH in $U(1)_{C}$ case. The contours show $\sigma_{\mathrm{prod}}$ (in ab/fb) at the $\sqrt s=14$ TeV LHC ($g_{C}=1$, solid) and at $\sqrt s=100$ TeV future collider ($g_{C}=0.3$, dashed) and ($g_{C}=1$, dot-dashed). Note : Blue region lies completely inside the green region.}
    \label{fig:1Cf}
\end{figure}

\begin{figure}
   \centering
    \includegraphics[width=0.49\textwidth]{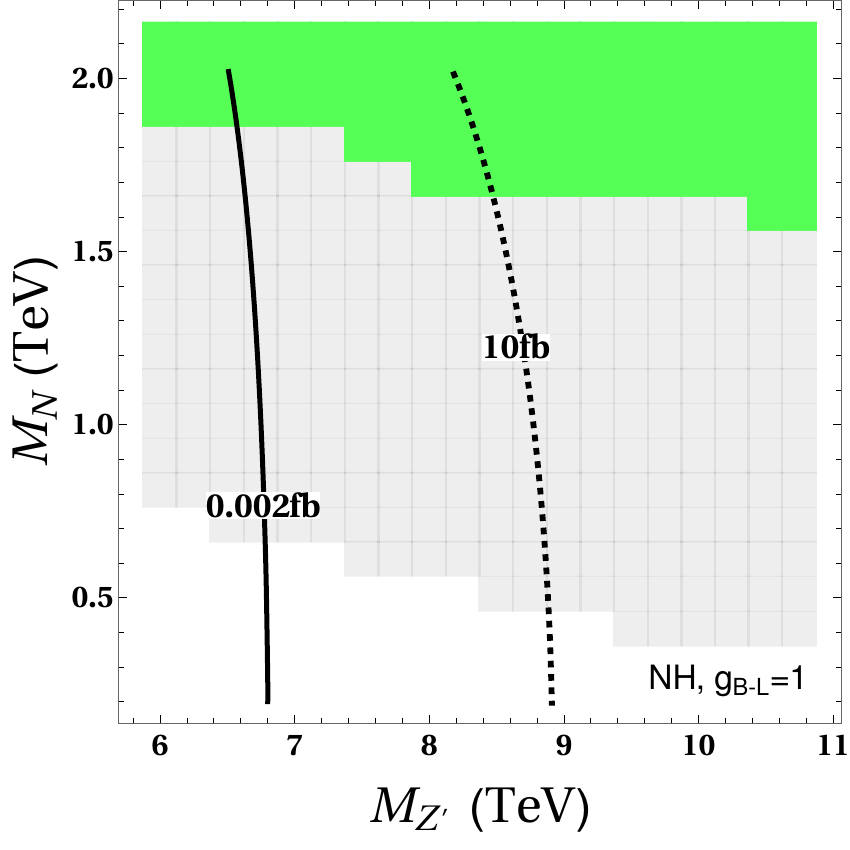}
    \includegraphics[width=0.49\textwidth]{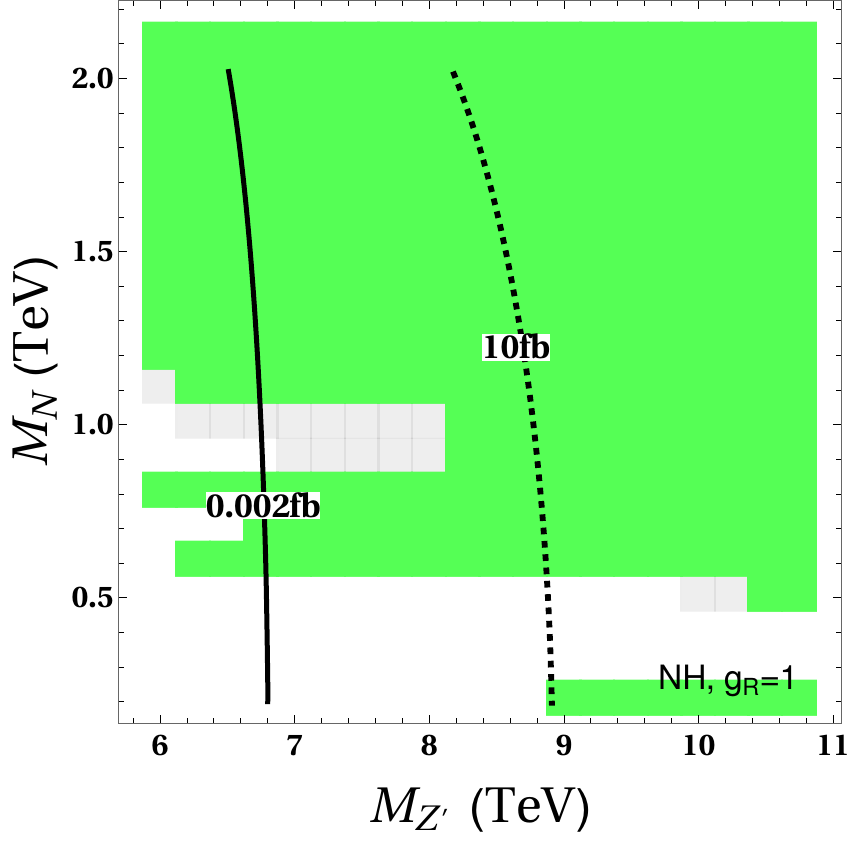}
    \\
    \includegraphics[width=0.49\textwidth]{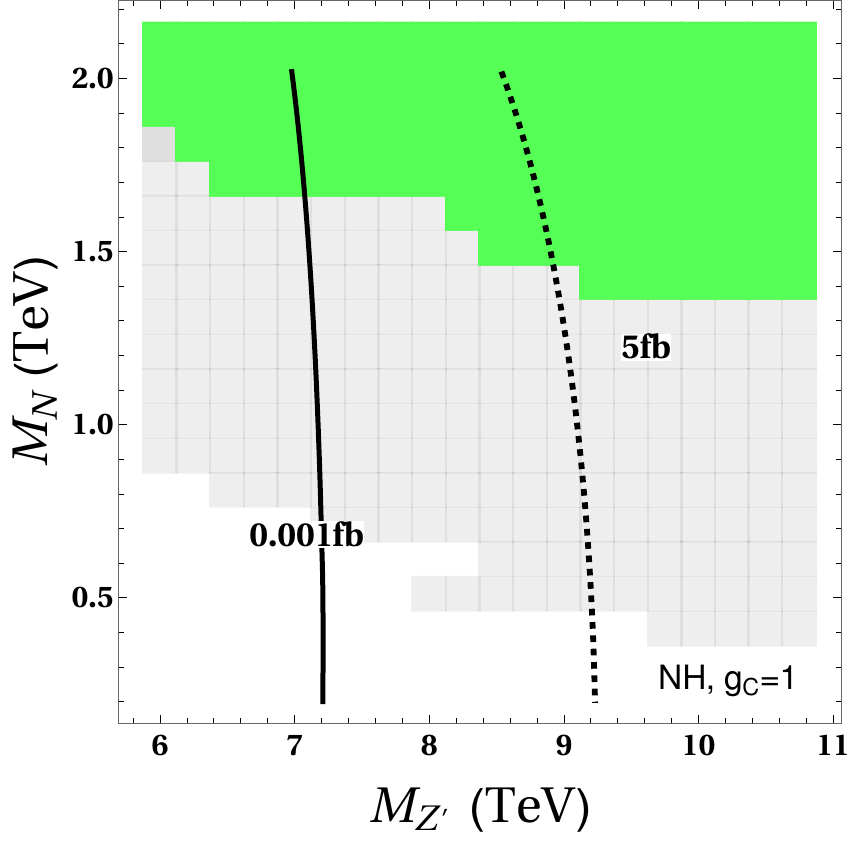}
    \caption{ Baryon asymmetry production in flavored case (unflavored) for $\eta_B \geq \eta_B^{\rm obs}$ shown in green (grey) region in the $(M_{Z'},M_N)$ plane for NH and different $g_{X}=1$, where $X = {(B-L,R,C)}$. The contours show $\sigma_{\mathrm{prod}}$ (in ab/fb) at the $\sqrt s=14$ TeV LHC (solid) and at $\sqrt s=100$ TeV future collider (dashed).}
    \label{fig:2all}
\end{figure}

%%%%%%%%%%%%%%%%%%%%%%%%%%%%%%%%%%%%%%%%%%%%%%%%%%%%%%%
%\section{Implications of neutrino oscillation data}
%\label{osc}
%%%%%%%%%%%%%%%%%%%%%%%%%%%%%%%%%%%%%%%%%%%%%%%%%%%%%%%
\section{Results and Discussion}
\label{sec:results}
In this section, we compute the baryon asymmetry production in three scenarios with different $U(1)_X$ charges (described in Sec.~\ref{sec:model}), with the flavored Boltzmann transport formalism as described in previous section \cite{BhupalDev:2014pfm}. Firstly, we show the effect of mass-splitting on the predicted baryon asymmetry in the $U(1)_{B-L}$ case as a function of $(M_{Z'},M_N)$ for $g_{B-L}=1\,(0.3)$ for NH (IH) scenario. For these chosen values of the gauge-couplings, we choose to restrict $M_{Z'}\geq 6$ TeV, to be consistent with the dilepton bounds from LEP-II \cite{Das:2021esm}. The colored regions indicate the parameter space for $\eta_B \geq \eta_B^{\rm obs}$. The green region indicates optimum RHN mass splitting $\Delta M_{N,\text{opt}}$ and blue region indicates $\Delta M_N = 10\,\Delta M_{N,\text{opt}}$. We clearly find that setting the $\Delta M_N$ to the optimum value gives the maximum parameter space consistent with the condition $\eta_B \geq \eta_B^{\rm obs}$. On increasing the $\Delta M_N$ by a factor of 10, we notice the parameter space reduces while still being a subset of the maximum parameter space allowed for optimal $\Delta M_N$. This is a general feature of the baryon asymmetry production for all $U(1)_X$ charges. Hence, we will hereafter set $\Delta M_N$ to the optimum value in all cases. Furthermore, we also include the collider sensitivity of the relevant parameter space in $(M_{Z'},M_N)$ plane. The contours show $\sigma_{\mathrm{prod}}$ (in ab/fb) at the $\sqrt s=14$ TeV LHC (solid/dashed) and at $\sqrt s=100$ TeV future collider (dot-dashed, dotted). 

We now plot the predicted baryon asymmetry $\eta_B$ in the $(M_{Z'},M_N)$ plane for a fixed $g_{X}=(0.3,1)$ for both NH and IH in the $U(1)_X, \, {X =(B\text{-}L, R, C)}$, shown in Figs.~\ref{fig:1Bf},\ref{fig:1Rf} and \ref{fig:1Cf} for optimal $\Delta M_N$. For $U(1)_{B\text{-}L}$ in Fig.~\ref{fig:1Bf}, successful leptogenesis is possible for $M_N>0.5(0.8)$ TeV for NH(IH) and $g_{B-L}=0.3$. Upon increasing the gauge coupling $g_{B-L}=1$, the parameter space shrinks due to increased washout effects and dilution. In this case, successful leptogenesis is only possible for $M_N>1.5(1.8)$ TeV for NH(IH), with $\sigma_{\mathrm{prod}}^{\text{NH}}=2$ ab at LHC  and $\sigma_{\mathrm{prod}}^{\text{IH}}=10$ fb at future collider.

For $U(1)_{R}$ in Fig.~\ref{fig:1Rf}, it can be seen that successful leptogenesis is possible for $M_N$ as low as $0.2$ TeV for both NH and IH if $g_{R}=0.3$, while for $g_{R}=1$, $M_N>0.2(1.2)$ TeV for NH(IH) is required. For all these cases, $\sigma_{\mathrm{prod}}$ at LHC reaches $\sim 1$-$1.5$ ab. Similarly for $U(1)_{C}$ in Fig.~\ref{fig:1Cf}, successful leptogenesis is possible for $M_N>0.4(0.6)$ TeV for NH(IH) and $g_{C}=0.3$, while for $g_{C}=1$, $M_N>1.3(1.6)$ TeV for NH(IH) is required. The $\sigma_{\mathrm{prod}}$ reach at LHC for NH and IH for $g_{C}=1$ is around $\sim 1$ ab. 

In addition to studying the flavored case, we also compare our results with the unflavored regime for all three $U(1)_X$ scenarios as shown in Fig.~\ref{fig:2all}. In this figure, we plot the prediction for the baryon asymmetry production in the flavored case (shown in red) and unflavored case (shown in grey) in the $(M_{Z'},M_N)$ plane. We have chosen to set the respective gauge coupling in each case to be unity and with ordering set to NH. It can be clearly seen that in all cases, the unflavored treatment usually overestimates the parameter space for required baryon asymmetry. For eg., successful leptogenesis in $U(1)_{B\text{-}L}$ for unflavored case is possible for $M_N>0.3$ TeV, while in the proper treatment involving the flavored transport equations, $M_N>1.5$ TeV is required. Hence, flavor effects play an important role in determination of $\eta_B$ generated through resonant leptogenesis \cite{Dev:2017trv}.

\section{Conclusion}
\label{conc}
We have studied the generation of baryon asymmetry through the resonant leptogenesis for the $U(1)_X$ extension of the SM. We numerically solve the flavored Boltzmann transport equations to calculate the total baryon asymmetry generated. After maximizing the $\eta_B$ for given $M_N$ and $M_{Z'}$, we show that the three different cases considered in this work can naturally explain the observed baryon asymmetry of the Universe in the large portion of the available parameter space. We find that the $U(1)_C$ case offers successful leptogenesis in a larger portion of the parameter space as compared to $U(1)_{B\text{-}L}$ and $U(1)_{R}$. We also perform a comparative study between the flavored and unflavored leptogenesis, showcasing the impact and importance of the flavor effects at play. Finally, we have also studied the collider prospects for all these different scenarios. We find that although HL-LHC might not be able to probe all the relevant parameter space, $\sqrt s=100$ TeV future collider can access these regions, thereby offering an opportunity to test resonant leptogenesis in the $U(1)_X$ model.  

\section*{Acknowledgements}\label{sec:acknowledgements}

GC thanks Arindam Das and Bhupal Dev for discussions and for collaboration during the early stages of this work. The work of GC is supported by the U.S. Department of Energy under the award number DE-SC0020250 and DE-SC0020262. GC also acknowledges the Center for Theoretical Underground Physics and Related Areas (CETUP* 2024) and the Institute for Underground Science at SURF for hospitality and for providing a stimulating environment, where this work was finalized.
%%%%%%%%%%%%%%%%%%

\medskip
\noindent 
{\bf Note Added:} While we were finalizing this work, Ref.~\cite{Das:2024gua} appeared, primarily based on an earlier version of this draft. But they have not included the flavor effects, which are known to be important for TeV-scale leptogenesis~\cite{Dev:2017trv}. Neither did they perform a scan of the $(M_N,M_{Z'})$ parameter space for collider tests as presented here. 
\appendix
\section{Appendix}
\label{append}
%%%%%%%%%%%%%%%%%%%
In this section, we provide the reduced cross sections used in this work for calculating the $\eta_B$ using flavored Boltzmaan transport equations. The form of $\hat{\sigma}= \frac{8\sigma}{s}[(p_{\text{in}}. p_{\text{out}})^2-m_f^4]$ where $\sigma$ is the total cross section in center of mass frame for the processes participating in the resonant leptogenesis, $p_{\text{in}}$ and $p_{\text{out}}$ are the momenta of incoming and outgoing state particles. The reduced cross sections involving Higgs $(h)$ have been obtained following \cite{Plumacher:1996kc}.
\begin{itemize}
\item[(a)] Scalar $(h)$ mediated $\ell N^i \to t q$ process in $s$-channel
\bea
\hat{\sigma}(N^i \ell \leftrightarrow t q)=\frac{3 \pi \alpha^2 m_t}{M_W^4 \sin\theta_W^4} (M_D^\dagger M_D)_{ij} \Bigg(1-\frac{m_{N_i}^2}{s}\Bigg)^2
\eea
Scalar $(h)$ mediated $ N^i t \to \ell q$ process in $t$-channel
\bea
\hat{\sigma}(N^j t \to \ell q)= \frac{3 \pi \alpha^2 m_t}{M_W^4 \sin\theta_W^4} (M_D^\dagger M_D)_{ij} \Bigg[1-\frac{m_{N_i}^2}{s}+\frac{m_{N_i}^2}{s}\log\Bigg(1+\frac{s-m_{N_i}^2}{m_h^2}\Bigg)\Bigg]
\eea
where $\alpha=\frac{e^2}{4 \pi}$.
\item[(b)] $\ell h \to \ell h$ process mediated by $N$ in the $s$-channel and $t$- channel
\bea
\hat{\sigma}(\ell h \leftrightarrow \ell h)&=& \frac{2\alpha^2 \pi~m_{N_1}^2}{ \sin^4\theta_W M_W^4 s} \Bigg\{\sum_{i=1}^{2} \frac{m_{N_{i}}^2}{m_{N_1}^2} (m_D^\dagger m_D)_{ii}^2\Bigg[\frac{s}{m_{N_i}^2}+ \frac{2 s}{m_{N_1}^2 \mathcal{D}_i}+ \nonumber \\
&&\frac{s^2}{2 m_{N_1}^4\mathcal{D}_i^2}-\Bigg(1+2\frac{s+m_{N_i}^2}{m_{N_1}^2\mathcal{D}_i}\Bigg)\log\Bigg(1+\frac{s}{m_{N_j}^2}\Bigg)\Bigg]+2 \frac{m_{N_2}}{m_{N_1}} \nonumber \\ 
&& \mathcal{R}e\Bigg[(m_D^\dagger m_D)_{12}^2)\Bigg] \Bigg[\frac{s}{m_{N_1}^2 \mathcal{D}_1}+\frac{s}{m_{N_1}^2 \mathcal{D}_2}+\frac{s^2}{2 m_{N_1}^4 \mathcal{D}_1\mathcal{D}_2}- \nonumber \\
&&\frac{(s+m_{N_1}^2)(s+m_{N_1}^2-2 m_{N_1}^2 m_{N_2}^2)}{ (m_{N_1}^2-m_{N_2}^2)\mathcal{D}_2} \log\Bigg(\frac{s}{m_{N_1}^2}+1\Bigg) \nonumber \\
&-& \frac{(s+m_{N_1}^2) (s+ m_{N_1}^2-2 m_{N_1} m_{N_2})}{(m_{N_1}-m_{N_2}) m_{N_1}^3\mathcal{D}_2 } \log\Big(\frac{s+m_{N_1}^2}{m_{N_1}^2}\Big)\Bigg] \nonumber \\
&-& \frac{(s+m_{N_1} m_{N_2}) (s+m_{N_1} m_{N_2} -2 m_{N_1}^2)}{(m_{N_1}-m_{N_2}) m_{N_1}^3\mathcal{D}_2} \log\Bigg(\frac{s+m_{N_1} m_{N_2} }{m_{N_1} m_{N_2}}\Bigg)\Bigg\}~~~
\eea
where $\mathcal{D}_i= \frac{(s-m_{N_j} m_{N_1})^2+ m_{N_j} m_{N_1} \Gamma_{ij}^2}{m_{N_1}^2 (s-m_{N_i} m_{N_1})}$ is the off-shell part of the propagator. 
\item[(c)] $\ell \ell \leftrightarrow h h $ process mediated by $N$ in the $t$- channel
\bea
\hat{\sigma}(\ell \ell \leftrightarrow h h)&=& \frac{2 \pi \alpha^2}{M_{W}^4 \sin\theta_W^4} \Bigg\{\sum_{i=1}^{2} \frac{m_{N_i}}{m_{N_1}}(m_D^\dagger m_D)_{ii}^2 \Bigg[ \frac{m_{N_1} s}{2 m_{N_j} (s+ m_{N_j} m_{N_1})} \nonumber \\
&+&\frac{m_{N_1}^2}{(s+ 2 m_{N_j} m_{N_1})} \log\Bigg(\frac{s+ m_{N_1} m_{N_j}}{m_{N_1} m_{N_j}}\Bigg)\Bigg]+\mathcal{R}e\Bigg[(m_D^\dagger m_D)_{12}^2\Bigg]\nonumber \\
&& \frac{\sqrt{m_{N_1} m_{N_2}} m_{N_1}^2}{(m_{N_1}- m_{N_2})(s+ m_{N_1}+ m_{N_1} m_{N_2})}\Bigg[  \frac{s+2 m_{N_1}^2}{m_{N_1}^2} \log\Bigg(\frac{s}{m_{N_1} m_{N_2}}+1\Bigg) \nonumber \\
&-& \frac{s+ 2 m_{N_2} m_{N_1}}{m_{N_1}^2} \log\Bigg(\frac{(s+m_{N_1}^2)}{m_{N_1}^2}\Bigg) 
\Bigg]
\Bigg\}.
\eea
\item[(d)] Pair production of $N$ from different initial SM charged fermions:
\bea
\hat{\sigma}(f \bar{f} \leftrightarrow N_i N_i )= \frac{g_X^4 x_\Phi^2 s^2}{12 \pi } \Bigg[\frac{(1-\frac{4 m_{N_i}^2}{s})^{\frac{3}{2}}}{(s-{m_Z^\prime}^2)^2+ {\Gamma_Z^\prime}^2 {m_Z^\prime}^2}\Bigg] (C_A^2+ C_V^2)
\label{RHN-1}
\eea
where the vector and axial-vector couplings are given in Tab.~\ref{tab-2} considering $x_\Phi=1$. This cross section should be averaged over the color factor for quarks.
\begin{table}[t]
\begin{center}
\begin{tabular}{|c|c|c|}
\hline
      Type of fermion  & vector coupling $(C_V)$& axial vector coupling $(C_A)$  \\ 
                       & $\frac{q_{f_L}^{}+q_{f_R}^{}}{2}$& $\frac{q_{f_L}^{}-q_{f_R}^{}}{2}$\\
\hline
charged lepton $(\ell)$ &$-\frac{3}{4}x_H^{}-1$&$\frac{1}{4} x_H^{}$\\
&&\\
%SM-like neutrinos $(\nu_L^\alpha)$&$\frac{1}{4} x_H^{}+\frac{1}{2}$&$\frac{1}{4} x_H^{}+\frac{1}
%{2}$\\
%&&\\
%\hline
%\hline
up-type quarks $(q_{u})$&$\frac{5}{12} x_H^{}+\frac{1}{3}$&$-\frac{1}{4} x_H^{}$\\
&&\\
down-type quarks $(q_{d})$&$-\frac{1}{12} x_H^{}+\frac{1}{3}$&$-\frac{1}{4} x_H^{}$\\
\hline
\end{tabular}
\end{center}
\caption{Vector and axial-vector couplings of different SM charged fermions with $Z^\prime$ where the axial vector couplings vanish for $x_H=0$ case. 
}
\label{tab-2}
\end{table}  
\item[(e)] For $i\neq j$, RHN pair production cross section is 
\bea
\sigma(N_i N_i \leftrightarrow N_j N_j )&=& \frac{g_X^4 x_\Phi^4\sqrt{(s-4 m_{N_j}^2)(s-4 m_{N_i}^2)}}{72 \pi \{(s-m_{Z^\prime}^2)^2+\Gamma_{Z^\prime}^2 m_{Z^\prime}^2\}} \nonumber \\
&& \Bigg[(s-4 m_{N_j}^2)(s-4 m_{N_i}^2)+ 12 \frac{m_{N_i}^2 m_{N_j}^2}{m_{Z^\prime}^4} (s- m_{Z^\prime})^2\Bigg]
\label{RHN-2}
\eea
For the other case, 
\bea
\hat{\sigma}(N_i N_i \leftrightarrow N_i N_i )&=& \frac{ g_X^4 x_\Phi^4}{128\pi}\Bigg[\frac{(s-4 m_{N_i}^2)^3}{3 s((s-m_{Z^\prime}^2)^2+m_{Z^\prime}^2 \Gamma_{Z^\prime}^2)}+ \frac{(s-4 m_{N_i}^2)}{s m_{Z^\prime}^4 (s-4 m_{N_i}^2+ m_{Z^\prime}^2)} \Big\{m_{Z^\prime}^2(s-4 m_{N_i}^2)^2+ \nonumber \\ 
&&2 (m_{Z^\prime}^2 -2 m_{N_i}^2)^3+ s(8 m_{N_i}^4+ 3 m_{Z^\prime}^4)-4 m_{N_i}^2 (m_{Z^\prime}^4+ 4 m_{N_i}^4+ m_{Z^\prime}^2 m_{N_i}^2)\Big\}+ \nonumber \\
&&+\Big\{\frac{(3 m_{Z^\prime}^2-4 m_{N_i}^2)(s-4 m_{N_i}^2)^2+ m_{Z^\prime}^4(3s-20 m_{N_i}^2)+ 2 m_{Z^\prime}^2 (m_{Z^\prime}^4+8 m_{N_i}^4)}{s m_{Z^\prime}^2 (s-4 m_{N_i}^2+ 2 m_{Z^\prime}^2)}\Big\}\nonumber \\ 
&& \log\Big(\frac{m_{Z^\prime}^2}{s-4 m_{N_i}^2+ m_{Z^\prime}^2}\Big)\Bigg]~~~~~
\eea
\item[(f)] Pair production of $N$ from Higgs $(h)$ in the $s$-channel:
\bea
\hat{\sigma}(h h \leftrightarrow N_i N_i)= \frac{{g_X^4} x_\Phi^4}{12 \pi} \frac{(x-4)^{\frac{3}{2}} (1-\frac{4 {m_\Phi}^2}{x m_{N_i}^2})^{\frac{3}{2}} \sqrt{x}}{c y+(x-y)^2}
\eea
where  $x=\frac{s}{m_{N_i}^2}$, $y=\frac{m_{Z^\prime}^2}{m_{N_i}^2}$, $c=\frac{\Gamma_{Z^\prime}^2}{m_{N}^2}$ and $\Gamma_{Z^\prime}$ is the total decay width of $Z^\prime$. Reduced cross section for $\Phi$ pair production from RHN in $t-$channel and $u-$channel processes:
\bea
\hat{\sigma}(N_i N_i \leftrightarrow h h)= \frac{Y_N^4}{8 \pi} \Bigg(1-\frac{4 m_{N_i}^2}{s}\Bigg)\Bigg[\frac{s-4m_{N_i}^2}{2 m_{N_i}^2}+\Bigg(1-\frac{4 m_{N_i}^2}{s}\Bigg)^{\frac{3}{2}}+ \log \Bigg(\frac{s-\sqrt{s(s-4 m_{N_i}^2)}}{s+\sqrt{s(s-4 m_{N_i}^2)}}\Bigg)\Bigg]~~~~~
\eea
\end{itemize}
%%%%%%%%%%%%%%%%
\bibliographystyle{JHEP}
\bibliography{ref}
%%%%%%%%%%%%%%%%%%%%%%%%%%%%%%%%%

\end{document}